\documentclass[prd,twocolumn,superscriptaddress,altaffilletter,showpacs,nofootinbib]{revtex4}
\usepackage[dvips]{graphicx}
\usepackage{amsmath}

\newcommand{\be}{\begin{equation}}
\newcommand{\ee}{\end{equation}}
\newcommand{\bea}{\begin{eqnarray}}
\newcommand{\eea}{\end{eqnarray}}

\begin{document}

\title{Power-Law NLED-Based Magnetic Universe Can Mimic Phantom Behavior}

\author{Ricardo Garc\'{\i}a-Salcedo}\email{rigarcias@ipn.mx}\affiliation{Centro de Investigacion en Ciencia Aplicada y Tecnologia Avanzada - Legaria
del IPN, M\'{e}xico D.F., M\'{e}xico.}

\author{Tame Gonzalez}\email{tame@uclv.edu.cu}\affiliation{Departamento de F\'{\i}sica, Universidad Central de Las Villas, 54830 Santa
Clara, Cuba.}

\author{Claudia Moreno}\email{claudia.moreno@cucei.udg.mx}\affiliation{Departamento de F\'{\i}sica y Matem\'aticas, Centro Universitario de
Ciencias Ex\'actas e Ingenier\'{\i}as, Corregidora 500 S.R., Universidad de Guadalajara, 44420 Guadalajara, Jalisco, M\'exico}

\author{Israel Quiros}\email{iquiros@fisica.ugto.mx}\affiliation{Divisi\'on de Ciencias e Ingenier\'ia de la Universidad de Guanajuato, A.P. 150, 37150, Le\'on, Guanajuato, M\'exico.}

\date{\today}

\begin{abstract}
We study the cosmic dynamics of a magnetic universe supported by non-linear electrodynamics (NLED) Lagrangeans that are proportional to powers of the electromagnetic invariant $\propto F^{1/(1-m)}$ ($m$ is an overall constant). For simplicity we focus in the case when $F$ depends on the magnetic field alone, a case dubbed in the bibliography as ''magnetic universe''. Our results demonstrate that, depending on the values of the free parameter $m$, the magnetic field can mimic phantom field behavior, an effect previously found in other contexts. It is demonstrated that, since there are found equilibrium points in the phase space of these models that can be associated with magnetic-dominated past and future attractors, a combination of positive and negative powers of $F$ may lead to interesting cosmological behavior. In particular, a cosmological scenario where the universe might evolve from a past NLED-driven (non-inflationary) state into a future (late-time) -- also NLED-driven -- inflationary stage, transiting through a matter-dominated solution, is envisioned. The impact of braneworld gravity on the dynamics driven by such NLED Lagrangeans is also investigated. It is demonstrated that, due to phantom property at late times, the non-linear electromagnetic effects may play an important role in deciding the fate of the cosmic evolution. Randall-Sundrum brane effects, in particular, modify the nature of the starting point of the cosmic evolution, as well as the fate of the expansion: both, the big bang singularity and the big rip event -- inherent in general relativity with phantom matter source --, may be avoided.
\end{abstract}

\maketitle

\section{Introduction}

Studying the equations of the non-linear electrodynamics (NLED) is an attractive subject of research in general relativity (GR) thanks to the fact that such quantum phenomena as vacuum polarization can be implemented in a classical model through their impact on the properties of the background space-time. Exact solutions of the Einstein's field equations coupled to NLED may hint at the relevance of the non-linear effects in strong gravitational and magnetic fields. It has been speculated, in particular, that very strong electromagnetic fields might help avoiding the occurrence of space-time singularities in the cosmological context \cite{27}. The impact of very strong electromagnetic fields (and of the NLED effects) regarding the causality issue in cosmology is also of relevance \cite{Novello0}. A different subject of research within the cosmological setting in GR coupled to NLED, is related with the chance for the NLED field to fuel primordial inflation. The cosmological inflationary scenario was proposed for the first time in the Ref. \cite{20} where, in order to overcome several problems of the standard model -- such as the flatness and the horizon problems, amongst others --, it was anticipated that a self-interacting scalar field with a particular form of the self-interaction potential as the source of the Einstein's field equations, might cause the universe to expand in an inflationary (super-accelerated) fashion. The inflationary paradigm has been supported by the observational evidence \cite{21}. In the reference \cite{15}, for instance, an anisotropic cosmological model coupled to Born-Infeld NLED was explored. It was found that this model might explain early time inflation \cite{28}. Thanks to the inflationary stage, the initially anisotropic universe might eventually isotropize. An alternative non-isotropic model where, additionally to the Born-Infeld NLED field, a cosmological constant was added to the Einstien's field equations, has been studied in \cite{Vollick0}. Yang-Mills cosmology with a non-Abelian Born-Infeld action has been also considered \cite{17}.

Magnetic universes (vanishing electric component) have been investigated within the context of NLED theories given by the Lagrangean density $L_1=-\frac{1}{4}F+\alpha F^2+\beta G^2$ (see below) \cite{16}. The non-linear term $\propto F^2$ may cause the universe to bounce thus avoiding the initial (big-bang) singularity. However, in Ref.\cite{Vollick1} it was demonstrated that the inclusion of a non-vanishing electric component $E$ where $E^2\simeq B^2$, removes the bounce and, in consequence, the universe starts its evolution in a singular state. Lagrangeans with inverse powers of the electromagnetic field $F$ are interesting because the non-linear electromagnetic effects might become important not only at early times in the cosmic evolution, but also at late times. Actually, models with Lagrangean density of the form $L_2=-\frac{1}{4}F-\frac{\gamma}{F}$ may account for the late-time stage of accelerated expansion of magnetic universes \cite{23}. In the later case, if a non-vanishing electric component $E$ is considered, accelerated expansion is allowed only when $E^2<3B^2$ \cite{Vollick1}. 

Even if the above -- very simplified -- models of non-linear electrodynamics coupled to general relativistic cosmology, describe hypothetical systems reminiscent of the fields in the real world, these models comprise interesting dynamical behavior that is worthy of independent investigation. Through the study of the equivalent phase spaces, for instance, it may be revealed how generic can be the occurrence of a bounce instead of the big-bang singularity, or, also, how the dynamics of these models is modified by the brane effects, etc. Aim of the present investigation is to explore the asymptotic properties of cosmological models where general relativity is coupled to NLED given by Lagrangeans that contain powers of the electromagnetic invariant $F$ ($L_1$ and $L_2$ above are particular cases). To simplify the study we shall consider magnetic universes (vanishing electric component). In spite of their simplicity these models contain interesting physics. We will rely on the use of the dynamical systems tools with the hope that such relevant concepts as past and/or future attractors, or saddle equilibrium points, could be correlated with generic cosmological behavior. Our goal will be to write the cosmological (Einstein's) equations in the form of an autonomous system of ordinary differential equations. Although these equations are non-linear, we will expand them in the neighborhood of the equilibrium (critical) points up to the linear approximation. Evaluating the sign of the real parts of the eigenvalues of the corresponding linearization matrices will allow us to judge about the stability of the critical points, which are correlated with asymptotic properties of the original system of equations. 

The paper has been organized in the following form. In section II we expose the basic mathematical and physical aspects of NLED-based cosmological models. The cosmic dynamics of models based on Lagrangeans that are power-law in the electromagnetic invariant $F$, is investigated in section III. For simplicity we concentrate in the so called ''magnetic universe'', where the electric component of the electromagnetic field is assumed vanishing. A concrete model, previously studied in \cite{23,novellocqg}, where positive and negative powers of $F$ are included in the NLED Lagrangean, is investigated in section IV. In section V we explore the impact of braneworld gravity on the dynamics of power-law NLED-magnetic universes. We include Randall-Sundrum, as well as Dvali-Gabadadze-Porrati brane models \cite{rs,dgp}. Section VI is dedicated to the physical discussion of the results obtained, and to the statement of brief conclusions. In order to make the paper self-contained, an appendix with brief tips on how to apply the dynamical systems tools to situations of cosmological interest is included. Here we use natural units ($8\pi G=8\pi/m_{Pl}^2=\hbar=c=1$).

\section{Nonlinear Electrodynamics coupled to General Relativity}

The four-dimensional (4D) Einstein-Hilbert action of gravity coupled to NLED is given by

\be S=\int d^4x\sqrt{-g}\left[ R+L_\gamma+L(F,G)\right],\label{action}\ee where $R$ is the curvature scalar, $L_\gamma$ -- the background perfect fluid's Lagrangean, and $L(F,G)$ is the gauge-invariant electromagnetic Lagrangean, which is a function of the electromagnetic invariants $F\equiv F^{\mu \nu}F_{\mu \nu }$ and $G\equiv\frac{1}{2}\epsilon_{\alpha\beta\mu\nu}F^{\alpha\beta}F_{\mu\nu}$ \cite{novellocqg}. As usual, the electromagnetic tensor is defined as: $F_{\mu\nu}=A_{\nu,\mu}-A_{\mu,\nu}$ (here the comma denotes partial derivative in respect to the spacetime coordinates, while the semicolon denotes covariant derivative instead). Standard (linear) Maxwell electrodynamics is given by the Lagrangean $L(F)=-F/4$. The corresponding field equations can be derived from the action (\ref{action}) by performing variations with respect to the spacetime metric $g_{\mu\nu}$, to obtain: $$G_{\mu\nu}=T_{\mu\nu}^\gamma+T_{\mu\nu}^{EM}\;,$$ where 

\bea &&T_{\mu\nu}^\gamma=\left(\rho_{\gamma}+p_{\gamma}\right) u_{\mu}u_{\nu}-p_{\gamma}g_{\mu\nu},\label{pf_t}\\
&&T_{\mu\nu}^{EM}=g_{\mu\nu}\,\left[L(F)-G L_G\right]-4F_{\mu\alpha}F_\nu^{\;\;\alpha}\,L_F,\label{em_t}\eea with $\rho_{\gamma}=\rho_{\gamma}(t)$, $p_\gamma=p_\gamma(t)$ -- the energy density and barotropic pressure of the background fluid, respectively, while $L_F\equiv dL/dF$, $L_{FF}\equiv d^2L/dF^2$, etc. Variation with respect to the components of the electromagnetic potential $A_\mu$ yields to the electromagnetic field equations 

\be \left(F^{\mu\nu}\,L_F+\frac{1}{2}\epsilon^{\alpha\beta\mu\nu}F_{\alpha\beta}L_G\right)_{;\mu}=0.\label{em-field}\ee

In this paper we shall consider a homogeneous and isotropic Friedmann-Robertson-Walker (FRW) metric of the form $$ds^{2}=dt^{2}-a(t)^{2}\left[\frac{dr^{2}}{1-kr^{2}}+r^{2}(d\theta^{2}+\sin ^{2}{\theta }d\phi ^{2})\right]\;,$$ where $a(t)$ is the cosmological scale factor, and the constant $k=-1,0,+1$ parametrizes the curvature of the spatial sections. In order to meet the requirements of homogeneous and isotropic cosmologies (as, in particular, the one associated with FRW spacetimes), the energy density and the pressure of the NLED field should be evaluated by averaging over volume. To do this, we define the volumetric spatial average of a quantity $X$ at the time $t$ by  (for details see \cite{novellocqg} and references therein): 

\be \overline{X}\equiv \lim_{V\rightarrow V_{0}}\frac{1}{V}\int d^3x \sqrt{-g}\;X,\ee where $V=\int d^3x\sqrt{-g}$ and $V_{0}$ is a sufficiently large time-dependent three-volume. Following the above averaging procedure, for the electromagnetic field to act as a source for the FRW model we need to impose that (the Latin indexes run over three-space);

\be \overline{E}_{i}=0,\;\;\;\overline{B}_{i}=0,\;\;\;\overline{E_{i}B_{j}}=0,\ee

\be \overline{E_{i}E_{j}}=-\frac{1}{3}E^{2}g_{ij},\;\;\;\overline{B_{i}B_{j}}=-\frac{1}{3}B^{2}g_{ij}.\ee Additionally it has to be assumed that the electric and magnetic fields, being random fields, have coherent lengths that are much shorter than the cosmological horizon scales. Under these assumptions the energy-momentum tensor of the electromagnetic (EM) field -- associated with the Lagrangean density $L=L(F,G)$ -- can be written in the form of the energy-momentum tensor for a perfect fluid \cite{novellocqg}:

\be T_{\mu\nu}^{EM}=\left(\rho_B+p_B\right) u_{\mu}u_{\nu}-p_Bg_{\mu\nu},\label{Tpf}\ee where

\bea &&\rho_B=-L+G L_G-4L_F E^{2},\label{rho} \\
&&p_B=L-G L_G-\frac{4}{3}\left( 2B^{2}-E^{2}\right) L_F,\label{p}\eea $E$ and $B$ being the averaged electric and magnetic fields, respectively. In what follows, to simplify the analysis, we shall consider a FRW universe with flat spatial sections ($k=0$), filled with a ''magnetic fluid'', i. e., the electric component $E$ will be assumed vanishing. Even this simplified picture can give important physical insights.

\section{Magnetic Universe Based on Power-Law NLED}

For the purposes of the present investigation we shall consider electromagnetic Lagrangean densities that depend on the invariant $F$ only. As
mentioned in \cite{novellocqg}, a particularly interesting case arises when only the average of the magnetic part $B$ is different from zero, leading to
the so called magnetic universe. This case turns out to be relevant in cosmology as long as the averaged electric field $E$ is screened by the
charged primordial plasma, while the magnetic field lines are frozen \cite{lemoine}.

In this simpler case, the cosmological equations can be written in the following form:

\bea &&3H^{2}=\rho_\gamma-L,\;\;2\dot{H}=-\gamma\rho_{\gamma}+\frac{4}{3}F L_F,\nonumber\\
&&\dot{\rho}_{\gamma}+3H\gamma\rho_{\gamma}=0,\;\;\dot{F}+4HF=0,\label{feqs}\eea where $H={\dot a}/a$ is the Hubble parameter, $\gamma$ is the barotropic index of the background's perfect fluid ($p_\gamma=\left(\gamma-1\right)\rho_\gamma$), while $$\rho_B=-L\;,\;\;p_{B}=L-\frac{4}{3}F L_F\;,$$ and, as already mentioned, we are considering $F=2B^2$. The system of cosmological equations (\ref{feqs}) is a closed system: there are 4 unknown field variables $\left( H,B,L,\rho \right)$, and 4 differential equations. However, as with any system of non-linear (second-order) differential equations, it is a very difficult (and perhaps unsuccessful) task to find exact solutions. In such cases the dynamical systems tools come to our rescue. These very simple tools give us the possibility to correlate such important concepts like past and future attractors (also saddle equilibrium points) in the phase space, with generic solutions to the set of equations (\ref{feqs}), without the need to analytically solve them.\footnote{For concise recipes of how to apply these tools to situations of cosmological interest we refer the reader to the section Appendix.} 

In order to be able to apply these tools to the case of interest in this paper, we introduce the following phase space variables:  $$x=\frac{-L}{3H^{2}}\;,\;\;y=\frac{F L_F}{3H^{2}}\;,$$ which allow to translate the original problem from the space of the field variables $\left( H,B,L,\rho \right)$ into the $(x,y)$-phase plane. After the above choice of variables, the system of equations (\ref{feqs}) can be written in the form of the following autonomous system of ordinary differential equations (ODE):

\bea &&x'=-2\left( x\frac{\dot{H}}{H^{2}}-2y\right),\notag\\
&&y'=-2y\left(\frac{\dot{H}}{H^{2}}+2-2\frac{my}{x}\right),\notag\\
&&\frac{\dot{H}}{H^{2}}=-\frac{3}{2}\gamma (1-x)+2y\label{ASL}\eea 
where we have defined the new parameter $m=L_{FF} L/L_F^2$, and the prime denotes derivative with respect to the new time variable $\tau\equiv\ln a$ -- properly the number of e-foldings. The above autonomous system of ODE is not, in general, a closed system of equations, unless the parameter $m$ is restricted to be a constant. This will be , precisely, the case we will be considering here. After making appropriate physical considerations, the phase space corresponding to the above autonomous system of ODE can be written as, $$\Psi =\left\{(x,y)\left\vert\; 0\leq x\leq 1,\;y\in R\right.\right\}\;.$$ 

Since we are considering constant values of the parameter $m=L_{FF} L/L_F^2$, then, after integrating the above expression, we obtain:

\be L=L_0\; F^{1/(1-m)},\;\;m\neq 1\label{L(F)}\ee where $L_0$ is an integration constant -- which has to be set equal to $-1/4$ in order to be able to recover the standard Lagrangean for Maxwell ED when $m=0$ --, and $$L=L_0\exp{(\lambda F)},\;\;m=1,$$ where $L_0$ and $\lambda$ are integration constants.

For the cases of interest $m=const$ the phase space variables $x$, $y$ are not independent variables since their ratio is a constant: $$\frac{x}{y}=-\frac{L}{F L_F}=m-1\;.$$ Hence the phase space is just a one-dimensional space, and the autonomous system of ODE (\ref{ASL}) reduces to just one independent equation ($m\neq 1$):

\be x'=-\left[\frac{3(m-1)\gamma+4}{m-1}\right]x(x-1).\label{ode'}\ee

The relevant magnitudes of observational interest can be given in terms of the variable $x$ and of the parameter $m$:

\begin{eqnarray*}
&&\Omega_B=x,\;\Omega_\gamma=1-x,\;\omega_B=\frac{3m+1}{3(1-m)},\\
&&q=-1+\frac{3\gamma}{2}-\left[\frac{3(m-1)\gamma+4}{2(m-1)}\right]x.
\end{eqnarray*} Note that the equation of state (EOS) parameter of the NLED component $\omega_B$, is the same value no matter what the critical point is. It depends only on the value of the overall constant $m$. For $m=0$ the NLED fluid behaves just like standard electromagnetic radiation with $\omega_B=1/3$ (in fact linear electrodynamics), while as $m\rightarrow\infty$, $\omega_B\rightarrow -1$: the NLED fluid behaves as a cosmological constant (quantum vacuum).

The equilibrium points corresponding to this case are obtained by requiring that in Eq. (\ref{ode'}) $x'=0$. Two equilibrium points are found:

\begin{itemize}

\item{The matter (background fluid)-dominated solution ($x=\Omega_B=0$):} $$\Omega_\gamma=1,\;\;q=-1+\frac{3\gamma}{2}\;.$$

\item{The NLED-dominated solution ($x=\Omega_B=1$):} $$\Omega_\gamma=0,\;\;q=-\frac{m+1}{m-1}$$

\end{itemize} Note that, for the case $m=1$ (exponential Lagrangean above), since $x/y=0$ always, only the matter-dominated equilibrium point is found. In this case the EOS parameter of the magnetic fluid is undefined. This case will not be considered anymore in this paper.

To explore the linear stability of these critical points $x=x_c$, we expand the equation (\ref{ode'}) in their neighborhood and keep terms linear in the small perturbation $\epsilon$. For the first equilibrium point $x=x_c=0$, we obtain the following equation for the perturbation: $$\epsilon'=k\epsilon,\;\;k\equiv\left[\frac{3(m-1)\gamma+4}{m-1}\right]\;,$$ while, for the second critical point ($x=x_c=1$): $$\epsilon'=-k(\epsilon+1)\;.$$ We see that, for positive $k>0$ the matter-dominated solution is unstable, while the NLED-dominated phase is stable. For $k<0$ these equilibrium points interchange their stability properties: the NLED-dominated phase is unstable while the matter-dominated solution is stable. Consider, for instance, background dust with $\gamma=1$. In this case $$k=3\left(\frac{m+1}{m-1}\right)\;,$$ so that, for $m>1$ the matter-dominated solution is the past attractor, while the NLED inflationary solution ($\omega_B<0$, $q<0$) is the late-time attractor. For $m<1$, on the contrary, the NLED non-inflationary solution ($\omega_B>0$, $q>0$) is the past attractor while the matter-dominated -- also non-inflationary ($q=1/2$) -- solution is the future attractor. 

One then expects that a combination of NLED terms proportional to positive and negative powers of $F$, would eventually allow for a cosmological scenario where the universe might evolve from a past NLED-driven (non-inflationary) state into a future (late-time), also NLED-driven, inflationary stage, transiting through a matter-dominated solution. The study of such a potentially interesting scenario will be the subject of the next section.

\section{Unified Description of Bouncing Cosmology Followed by Late-Time Accelerated Expansion}

In this section we focus in the study of the asymptotic properties of a cosmological model with interesting features, namely a phase of current
cosmic acceleration and the absence of an initial singularity, which was proposed in \cite{novellocqg} (see also \cite{23}) and is based upon the
following Lagrangean density:

\be L=-\frac{1}{4}F+\alpha F^2+\beta F^{-1},  \label{L}\ee where $\alpha$ and $\beta$ are arbitrary (constant) parameters. As seen this Lagranagian contains both positive and negative powers of $F$. The second (quadratic) term dominates during very early epochs of the cosmic dynamics, while the Maxwell term (first term above) dominates in the radiation era. The last term in (\ref{L}) is responsible for the accelerated phase of the cosmic evolution \cite{novellocqg}. The above Lagrangean density yields a unified scenario to describe both the acceleration of the universe (for weak fields) and the avoidance of the initial singularity, as a consequence of its properties in the strong-field regime.

Recalling that we are considering magnetic universes, i.e., $F=2B^2$, where $B^2$ is an averaged value of the magnetic field,\footnote{For details of the averaging procedure consult \cite{novellocqg}. See also section II of the present paper.} the stress-energy tensor associated with (\ref{L}) can be written in the form of an equivalent perfect fluid stress-energy tensor with energy density and parametric pressure (see section II):

\bea &&\rho_B=\frac{B^{2}}{2}\left( 1-8\alpha B^{2}-\frac{\beta}{B^4}\right),\notag\\
&&p_B=\frac{B^2}{6}\left(1-40\alpha B^2+\frac{7\beta}{B^4}\right),\label{rhop}\eea respectively. Notice that, for large values of the NLED field, positivity of energy requires that $B<1/\sqrt{8\alpha }$, while, for small enough values of $B\ll 1$, if one considers positive $\beta>0$, then positivity of energy implies that $B>(7\beta)^{1/4}$. In the latter case, in the unified model given by the Lagrangean density (\ref{L}) \cite{novellocqg}, there arise both higher and lower bounds on the values the NLED field $B$ can take. The existence of the lower bound, at first sight might appear problematic, however, given that the observational data constraints the parameter $\beta$ to be $\sqrt{|\beta|}\approx 4\times 10^{-28}\;g\;cm^{-3}$ \cite{23}, then the lower bound on $B$ can be admitted without going into conflicts with observations.

In the rest of this section we assume the background fluid to be dust cold dark matter (CDM), so that $\gamma=1$. Our goal in this section will be to put the corresponding cosmological equations (\ref{feqs}): 

\begin{eqnarray} &&3H^2=\rho_{cdm}+\rho_B,\nonumber\\
&&2\dot H=-\rho_{cdm}-(\rho_B+p_B),\nonumber\\
&&\dot\rho_{cdm}+3H\rho_{cdm}=0,\nonumber\\
&&\dot\rho_B+3H(\rho_B+p_B)=0\;\Rightarrow\;\dot B=-2HB,\label{feqs cdm}\end{eqnarray} with $\rho_B$ and $p_B$ given by (\ref{rhop}), in the form of an autonomous system of ODE. In the present case, since there exists a combination of powers of $F$ in the Lagrangean, we can not use the variables of the former section. We choose the following phase space variables instead:

\be x\equiv\frac{\rho_B}{3H^2},\;\;y\equiv\frac{16\alpha B^4}{3H^2},\;\;z\equiv\frac{4\beta}{3H^2 B^2}.\label{xyz}\ee In terms of these variables one has: $$\frac{B^2}{6H^2}=x+\frac{y}{4}+\frac{z}{8},\;\;\frac{\rho_B+p_B}{H^2}=4x-y+z\;,$$ and, also $$2\frac{H'}{H}=-3-x+y-z\;,$$ meanwhile, the dimensionless energy density of the CDM can be expressed as function of the $x$-variable alone ($\Omega_B=x$);

\be\Omega_{cdm}\equiv\frac{\rho_{cdm}}{3H^2}=1-x.\label{omega}\ee Other magnitudes of observational relevance are the equation of state (EOS) parameter of the magnetic field, and the deceleration parameter: $$\omega_B=\frac{x-y+z}{3x},\;\;q=\frac{1+x-y+z}{2}\;,$$ respectively.

The following autonomous system of ODE is obtained out of (\ref{feqs cdm}):

\bea &&x'=(x-1)(x-y+z),\nonumber\\
&&y'=-y(5-x+y-z),\nonumber\\
&&z'=z(7+x-y+z).\label{asode}\eea

Here, for generality of the analysis we shall consider arbitrary $\alpha\in\Re$ and $\beta\in\Re$. The phase space relevant to the present study is then given by the following region in $(x,y,z)$:

\bea &&\Psi_U=\{(x,y,z)|0\leq x\leq 1,\;(y,z)\in\Re^2,\nonumber\\
&&\;\;\;\;\;\;\;\;\;\;\;\;\;\;\;\;\;\;\;\;\;\;\;\;\;\;\;\;\;\;\;\;\;\;\;8x+2y+z\geq 0\},\label{phi}\eea where we have taken into consideration that $$\frac{B^2}{6H^2}=x+\frac{y}{4}+\frac{z}{8}.$$

The equilibrium points of (\ref{asode}) in $\Psi_U$, their existence and stability properties, are discussed below:

\begin{enumerate}

\item{Radiation-dominated phase:} $$P_{rad}=(x,y,z)=(1,0,0),\;\;\Omega_B=1,\;\;\Omega_{cdm}=0\;.$$ This is a decelerating expansion solution ($q=1$), that is fueled by standard radiation with $\omega_B=1/3$. The eigenvalues of the linearization matrix corresponding to this equilibrium point are: $$\lambda_1=1,\;\;\lambda_2=-4,\;\;\lambda_3=8\;,$$ so that it is a saddle in $\Psi_U$.

\item{Infra-red NLED-dominated solution:} $$P_{nled}^{IR}=(1,0,-8),\;\;\Omega_B=1,\;\;\Omega_{cdm}=0\;.$$ This is a late-time, super-inflationary solution ($q=-3$), where the NLED fluid mimics phantom behavior ($\omega_B=-7/3$). This solution exist only for $z<0$ (negative $\beta<0$). In this case there is no lower bound on the magnitude of the magnetic field. The eigenvalues of the linearization matrix for $P_{nled}^{IR}$ are $$\lambda_1=-8,\;\;\lambda_2=-7,\;\;\lambda_3=-12\;,$$ so that this solution corresponds to a late-time (future) attractor. Since, this point is dominated by the NLED magnetic field mimicking phantom field, then, the late-time attractor could be associated with a cosmological singularity (most probably a big-rip type of singularity).

\item{Ultra-violet NLED-dominated phase:} $$P_{nled}^{UV}=(1,-4,0),\;\;\Omega_B=1,\;\;\Omega_{cdm}=0\;.$$ This solution corresponds to an early-time, super-stiff-fluid solution ($\omega_B=5/3$), which is associated with super-decelerating expansion ($q=3$). This solution exist only for $y<0$, i. e., if the constant $\alpha$ is a negative quantity ($\alpha<0$). Curiously, this case does not meet the conditions for a bounce (there is no upper bound on the magnitude of the magnetic field), and the corresponding cosmology starts with a big-bang singularity, since $B^2/H^2=0,\,B\neq 0,\;\Rightarrow\;H\rightarrow\infty$. This solution is a past attractor in $\Psi_U$, since the eigenvalues of the corresponding linearization matrix: $$\lambda_1=5,\;\;\lambda_2=12,\;\;\lambda_3=4\;,$$ are all positive quantities. This means that this point is the starting point of every probe path in the phase space $\Psi_U$.

\item{CDM-dominated solution:} $$P_{cdm}=(0,0,0),\;\;\Omega_B=0,\;\;\Omega_{cdm}=1\;.$$ This phase of the cosmic evolution is characterized by decelerated expansion ($q=1/2$). The EOS parameter for the magnetic field is undefined in this case. The existence of this solution is necessary for the formation of the observed amount of structure. This is also a saddle equilibrium point in $\Psi_U$, since: $$\lambda_1=-5,\;\;\lambda_2=7,\;\;\lambda_3=-1\;.$$

\end{enumerate} For positive definite $\alpha$ and $\beta$, only the radiation-dominated equilibrium point $P_{nled}$, and the CDM-dominated solution $P_{cdm}$, are found in $\Psi_U$.

The above results confirm our expectation that a combination of positive and negative powers of the electromagnetic invariant $F$ in the NLED Lagrangean, can drive a very interesting cosmological scenario leading to accelerated expansion at late times. Unfortunately, early-time (primordial) inflation can not be obtained in the present model. It is remarkable also that the bounce is not a generic solution in such models.

In the next section we shall explore the influence of the brane effects on the dynamics of the NLED-magnetic universe.

\section{NLED on the brane}

In this section we explore the possible effect of braneworld gravity on the above picture. A Randall-Sundrum braneworld model \cite{rs} has proved satisfactory to modify the primordial inflation scenario, since the extra-dimensional brane effects are appreciable at high-energies (large brane tension). A Dvali-Gabadadze-Porrati braneworld model, on the contrary, modifies gravity at large (cosmological) scales \cite{dgp}. Hence, the study of the cosmological dynamics driven by NLED-magnetic universes trapped in either RS or DGP braneworlds, is an interesting possibility to look for alternative, viable, cosmological scenarios.

\subsection{RS2 brane}

One of the most appealing braneworld scenarios is the Randall-Sundrum brane model of type 2 (RS2) \cite{rs}. In this model a single co-dimension 1 brane with positive tension is embedded in a five-dimensional anti-de Sitter (AdS) space�time, which is infinite in the direction perpendicular to the brane. In general, the standard model particles are confined to the brane, meanwhile gravitation can propagate in the bulk. In the low-energy limit, due to the curvature of the bulk, the graviton is confined to the brane, and standard (four-dimensional) general relativity laws are recovered.

The FRW cosmological equations for a RS2 brane with CDM and a perfect fluid of NLED trapped on it, can be written as;

\bea &&3H^2=\rho_T\left(1+\frac{\rho_T}{2\lambda}\right),\;\;\rho_T=\rho_{cdm}+\rho_B,\nonumber\\
&&2\dot H=-\left[\rho_{cdm}-\frac{4}{3}FL_{F}\right]\left(1+\frac{\rho_T}{\lambda}\right),\nonumber\\
&&\dot\rho_{cdm}=-3H\rho_{cdm},\;\;\dot F+4H F=0,\label{feqs'}\eea where $\lambda$ is the brane tension.

In order to write the above system of equations into the form of an autonomous system of ODE, we introduce the following phase space variables (recall that $\rho_B=-L$);
 
\be x=\frac{-L}{3H^2},\;\;y=\frac{FL_F}{3H^2},\;\;v=\frac{\rho_T}{3H^2}.\label{brane xyv}\ee After this we can write, $$\frac{\rho_T}{\lambda}=\frac{2(1-v)}{v},\;\;\Omega_{cdm}=v-x\;.$$ Notice that, in the above phase space variables, the high-energy regime of the brane is associated with the limit $v\rightarrow 0$, while the low-energy, general relativity limit corresponds to $v=1$.

The following autonomous system of ODE is found:

\bea &&x'=2\left(2y-x\frac{\dot H}{H^2}\right),\nonumber\\
&&y'=-2y\left(2-2\frac{m y}{x}+\frac{\dot H}{H^2}\right),\nonumber\\
&&v'=-3\left(v-x\right)+\frac{4}{3}y-2v\frac{\dot H}{H^2},\label{asode'}\eea where $$\frac{\dot H}{H^2}=-\frac{3(2-v)}{2v}\left[v-x-4y\right]\;,$$ and, as before, for $m\neq 1$ (see section III), $$m=\frac{L L_{FF}}{L_F^2}=const.\;\Rightarrow\;L=L_0 F^{1/(1-m)}\;.$$ Besides, since $y=x/(m-1)$, only one of these variables is independent, say, $x$. This means that, in fact, the system of ODE (\ref{asode'}) is a two-dimensional system of ODE:

\bea &&x'=x\left(\frac{4}{m-1}-2\frac{\dot H}{H^2}\right),\nonumber\\
&&v'=-3(v-x)+\frac{4x}{3(m-1)}-2v\frac{\dot H}{H^2},\label{asode''}\eea with $$2\frac{\dot H}{H^2}=-\frac{3(2-v)}{v}\left[v-\left(\frac{m+3}{m-1}\right)x\right]\;.$$ The phase space where to look for the equilibrium points of the above autonomous system of ODE is given by the following compact $(x,v)$-region: $$\Psi_{brane}=\{(x,v)|0\leq x\leq 1,\;0\leq v\leq 1\}\;.$$ 

The equilibrium points of (\ref{asode''}), their existence and stability properties are discussed below. Only two equilibrium points can be found:

\begin{itemize}

\item{CDM-dominated (GR) critical point:}  $$P_{cdm}^{GR}=(x,v)=(0,1),\;\;\Omega_{cdm}=1,\;\;\Omega_B=0\;.$$ This point is associated with decelerated expansion $q=1/2$. The eigenvalues of the corresponding linearization matrix are: $$\lambda_1=-3,\;\;\lambda_2=\frac{3m+1}{m-1}\;.$$ For $$-\frac{1}{3}<m<1\;,$$ the point $P_{cdm}^{GR}\in\Psi_{brane}$ is the late-time attractor, while for $$m<-\frac{1}{3}\;,\;\text{or},\;\;m>1\;,$$ it is a saddle in $\Psi_{brane}$.

\item{Scaling (RS) critical point:} $$P_{sc}^{RS}=(3k v_0,v_0),\;\;\Omega_{cdm}=(1-3k)v_0,\;\;\Omega_B=3k v_0\;,$$ where $$k\equiv\left(\frac{3m+1}{9m-5}\right),\;\;v_0\equiv\frac{57m+11}{3(11m+1)}\;.$$ The scaling character is due to the existence of non-vanishing ratio $$\frac{\Omega_{cdm}}{\Omega_B}=\frac{1-3k}{3k}=-\frac{8}{9m+3}\;.$$ This solution exists only for $m\leq-1/3$. The deceleration parameter is given by: $$q=-\frac{m+1}{m-1}\;.$$ The eigenvalues of the Jacobian matrix corresponding to this critical point are: $$\lambda_1=-\frac{3m+1}{m-1},\;\;\lambda_2=\frac{4(57m+11)}{(9m-5)(m-1)}\;.$$ It can be verified that, for the $m$-interval where the point exists, both eigenvalues are negative quantities, so that the scaling solution is always the late-time attractor. 

\end{itemize} 

Note that for $m>-1/3$, only one equilibrium point (the CDM-dominated solution) exists, while for $m\leq-1/3$ both critical points exist. In the latter case, the CDM-dominated solution is a saddle critical point in $\Psi_{brane}$, while the scaling solution is the late-time attractor. The latter attractor is associated with accelerated expansion whenever $m<-1$, while for $-1<m\leq -1/3$, the expansion is decelerated instead. Notice, also, that the NLED field EOS parameter $\omega_B:=-1-4/3(m-1)$, does not depend on $x$, $z$, so that is the same value for both the CDM-dominated solution and for the scaling one.

If we compare the above results with the ones of section III, we see that the RS2 brane does really modify the dynamics of the NLED-magnetic universe. Actually, the NLED-dominated solution reported in section III has been replaced by the scaling solution above. Hence, it makes sense to inspect the model of section IV when the brane effects are included.

\subsubsection{Combination of positive and negative powers of $F$}

Here we shall consider the Lagrangean (\ref{L}) which consists of a combination of positive and negative powers of the electromagnetic invariant $F$. The cosmological equations are the equations (\ref{feqs'}), which can be rewritten in a less formal way:

\bea &&3H^2=\rho_T\left(1+\frac{\rho_T}{2\lambda}\right),\;\;\rho_T=\rho_{cdm}+\rho_B,\nonumber\\
&&2\dot H=-(\rho_{cdm}+\rho_B+p_B)\left(1+\frac{\rho_T}{\lambda}\right),\nonumber\\
&&\dot\rho_{cdm}=-3H\rho_{cdm},\;\;\dot B=-2H B,\label{feqs''}\eea where $\rho_B$ and $p_B$ are defined in equations (\ref{rhop}). It is convenient to introduce the following phase space variables:

\be x\equiv\frac{\rho_B}{3H^2},\;\;y\equiv\frac{16\alpha B^4}{3H^2},\;\;z\equiv\frac{4\beta}{3H^2 B^2},\;\;v\equiv\frac{\rho_T}{3H^2}.\label{xyzv}\ee In terms of these variables one has, for instance: $$\frac{\rho_T}{\lambda}=2\left(\frac{1-v}{v}\right),\;\;\Omega_{cdm}=v-x\;,$$ and, also, $$2\frac{H'}{H}=-(3v+x-y+z)\left(\frac{2-v}{v}\right)\;.$$ The variable $v$ controls the brane regime, so that, for instance, the GR-limit (formal limit $\lambda\rightarrow\infty$) corresponds to $v=1$. The following autonomous system of ODE can be derived out of (\ref{feqs''}):

\bea &&x'=3(1-v)x+\left[\left(\frac{2-v}{v}\right)x-1\right](x-y+z),\nonumber\\
&&y'=y\left[-5+(1-v)+\left(\frac{2-v}{v}\right)(x-y+z)\right],\nonumber\\
&&z'=z\left[7+(1-v)+\left(\frac{2-v}{v}\right)(x-y+z)\right],\nonumber\\
&&v'=(1-v)(3v+x-y+z).\label{brane asode}\eea At $v=1$ the first three equations above coincide with the equations (\ref{asode}) of section IV, which hold for general relativity with a NLED-magnetic field. The phase space of the model can be defined as follows:

\bea &&\Psi_U^{brane}=\{(x,y,z,v)|0\leq x\leq 1,\;\;(y,z)\in\Re^2,\nonumber\\
&&\;\;\;\;\;\;\;\;\;\;\;\;\;\;\;\;\;\;\;\;\;\;\;\;8x+2y+z\geq 0,\;\;0\leq v\leq 1\}.\label{brane U}\eea 

For the following discussion it will useful to write several magnitudes of observational interest in terms of the above phase space variables ($\Omega_B=x$): $$\omega_B=\frac{x-y+z}{3x},\;\;q=\frac{4-3v}{2}+\left(\frac{2-v}{2v}\right)(x-y+z)\;.$$

The critical points of the autonomous system of ODE (\ref{brane asode}) in the phase space $\Psi_U^{brane}$, their physical properties and stability are discussed below. As in the general relativity case (section IV), four equilibrium points are found:

\begin{enumerate}

\item{CDM-dominated solution:} $$P_{cdm}=(0,0,0,1),\;\;\Omega_B=0,\;\;\Omega_{cdm}=1\;.$$ Since $q=1/2$, this solution is associated with decelerated expansion. The NLED-magnetic field EOS parameter $\omega_B$ is undefined. This is a saddle equilibrium point in $\Psi_U^{brane}$. Actually, the eigenvalues of the linearization matrix corresponding to this point are: $$\lambda_1=-5,\;\;\lambda_2=7,\;\;\lambda_3=-3,\;\;\lambda_4=-1\;.$$

\item{Radiation-dominated solution:} $$P_{rad}=(1,0,0,1),\;\;\Omega_B=1,\;\;\Omega_{cdm}=0\;.$$ It is also a decelerated-expansion solution ($q=1$) driven by standard (Maxwell) radiation ($\omega_B=1/3$). As the CDM-dominated phase, this solution also represents a saddle equilibrium point in $\Psi_U^{brane}$, since the eigenvalues of the linearization matrix are of opposite signs: $$\lambda_1=8,\;\;\lambda_2=1,\;\;\lambda_{3,4}=-4\;.$$

\item{UV NLED-dominated solution:} $$P_{nled}^{UV}=(1,-4,0,1),\;\;\Omega_B=1,\;\;\Omega_{cdm}=0\;.$$ This solution shares many properties with its similar in section IV: it is a super-decelerated ($q=3$), super-stiff state ($\omega_B=5/3$), associated with a big-bang-type singularity ($H\rightarrow\infty$). The new feature here is that the stability properties have been modified by the brane effects. Actually, the eigenvalues of the linearization matrix in the present case are: $$\lambda_1=12,\;\;\lambda_2=4,\;\;\lambda_3=5,\;\;\lambda_4=-8\;,$$ so that this is a saddle equilibrium point in $\Psi_U^{brane}$ (its similar in section IV is a past attractor).

\item{IR NLED-dominated solution:} $$P_{nled}^{IR}=(1,0,-8,1),\;\;\Omega_B=1,\;\;\Omega_{cdm}=0\;.$$ As in the former case, this solution shares many properties with its similar in section IV: it is a super-accelerated ($q=-3$), phantom-like solution ($\omega_B=-7/3$), possibly associated with a big-rip-type singularity. Other properties, the stability in particular, have been modified by the brane effects. Actually, the eigenvalues of the Jacobian matrix are $$\lambda_1=-12,\;\;\lambda_2=-8,\;\;\lambda_3=-7,\;\;\lambda_4=4\;,$$ so that it is also a saddle equilibrium point in the phase space (\ref{brane U}).

\end{enumerate} 

The first thing that is worthy of mention, is the fact that all of the above critical points represent saddle equilibrium points in the phase space $\Psi_U^{brane}$ (\ref{brane U}). Note that, since $v=1$, the four equilibrium points are associated with GR. In fact these coincide in almost all aspects with the ones found in section IV. The first two points $P_{cdm}$, and $P_{rad}$, show no fundamental differences with their similar in that section. However, the NLED-dominated solutions $P_{nled}^{UV}$ and $P_{nled}^{IR}$, have different stability properties than their corresponding solutions for the GR case: while in the latter case $P_{nled}^{UV}$ was the past attractor and $P_{nled}^{IR}$ was the future attractor, in the present case both are saddle equilibrium points as already said. This means, in turn, that the space-time singularities associated with these critical points (big-bang and big-rip singularities respectively), might be evaded in the present case. Modification of the stability properties of the UV solution was expected since RS brane effects are appreciable at early times (high energies). The surprise was the IR solution since, it is expected that at low energies/large cosmological scales, the RS2 brane effects can be safely ignored. We postpone the discussion of the possible origin of this effect to section VI.

\subsection{DGP brane}

Another braneworld scenario that has received much attention in the last years is the Dvali-Gabadadze-Porrati (DGP) model \cite{dgp}. It describes a brane with four-dimensional (4D) world-volume, that is embedded into a flat 5D bulk, and allows for infrared (IR)/large-scale modifications of gravitational laws. A distinctive ingredient of the model is the induced Einstein-Hilbert action on the brane, that is responsible for the recovery of
4D Einstein gravity at moderate scales, even if the mechanism of this recovery is rather nontrivial \cite{deffayet}. The acceleration of the expansion at late times is explained here as a consequence of the leakage of gravity into the bulk at large (cosmological) scales, so it is just a 5D geometrical effect. As with many IR modifications of gravity, there are ghosts modes in the spectrum of the theory \cite{ghosts}.\footnote{In fact there are ghosts only in one of the branches of the DGP model; the so-called ��self-accelerating�� branch, or self-accelerating cosmological phase \cite{ghosts'}. The Minkowski cosmological phase is free of ghosts.} Nevertheless, studying the dynamics of DGP models continues to be a very attractive subject of research. 

The cosmological FRW equations for a DGP brane with CDM and NLED-magnetic fluid trapped on it are the following:

\bea &&3Q_\pm^2=\left(\rho_{cdm}+\rho_B\right),\;\;Q_\pm^2\equiv H^2\pm\frac{H}{r_c}\nonumber\\
&&\frac{\dot Q}{Q}=-\frac{1}{2}H\left(\rho_{cdm}-\frac{4}{3} F L_F\right),\nonumber\\
&&\dot\rho_m=-3H\rho_{cdm},\;\;\dot F+4H F=0.\label{dgp feqs}\eea In order to write (\ref{dgp feqs}) as an autonomous system of ODE we introduce the following phase space variables:

\be x=-\frac{L}{3Q^2},\;\;v=\frac{Q}{H}.\label{xv}\ee The following system of ODE is found:

\bea &&x'=\left(\frac{3m+1}{1-m}\right)(x-1)x,\nonumber\\
&&v'=-\frac{v}{2}\left(\frac{1-v^2}{1+v^2}\right)\left[3+\left(\frac{3m+1}{1-m}\right)x\right].\label{dgp asode}\eea In terms of $x$, $v$, the relevant parameters can be written as: $$\Omega_{cdm}=v^2(1-x),\;\;\Omega_B=v^2x,\;\;\omega_B=\frac{m+1}{1-m}\;,$$ and the deceleration parameter $$q=-1-\frac{2}{3}\left[\frac{(3m+1)x-3(m-1)}{m-1}\right]\left(\frac{v^2}{1+v^2}\right)\;.$$ Note that the EOS parameter of the NLED-magnetic fluid $\omega_B$ depends only on $m$ and not on the concrete critical point being considered. For $m>1$ the NLED component mimics phantom fluid ($\omega_B<-1$).

According to which branch of the DGP is being considered, the phase space for this case is split into two regions that share a common subspace defined by the value $v=1$. This subspace corresponds to general relativity. For the self-accelerating branch (minus sign in the definition of $Q$) the phase space is given by

\be\Psi_{dgp}^-=\{(x,v)|0\leq x\leq 1,\;\;0\leq v\leq 1\},\label{self-a}\ee while, for the Minkowski cosmological phase (plus sign in the definition of $Q$) it is given by 

\be\Psi_{dgp}^+=\{(x,v)|0\leq x\leq 1,\;\;v\geq 1\}.\label{minkowski}\ee Points with negative $v$-s ($v<0$), are not considered due to the symmetry of the field equations (\ref{dgp feqs}) under the change $Q\rightarrow-Q$.

There are found two equilibrium points belonging in the common GR-boundary $\Psi_{dgp}^-\cap\Psi_{dgp}^+$:

\begin{itemize}

\item{CDM-dominated solution:} $$P_{cdm}^{GR}=(0,1),\;\;\Omega_{cdm}=1,\;\;\Omega_B=0,\;\;q=\frac{1}{2}\;.$$ Since the eigenvalues of the Jacobian matrix are $$\lambda_1=\frac{3}{2},\,\;\lambda_2=\frac{3m+1}{m-1}\;,$$ then, this point is a saddle ($\lambda_2<0$) for $m<-1/3$ or $m>1$. For $-1/3<m<1$, it is a source critical point/past attractor in $\Psi_{dgp}^-\cap\Psi_{dgp}^+$ instead.

\item{NLED-dominated solution:} $$P_{nled}^{GR}=(1,1),\;\;\Omega_{cdm}=0,\;\;\Omega_B=1,\;\;q=-\frac{m+1}{m-1}\;.$$ This solution is inflationary ($q<0$) whenever $|m|>1$. It is the late-time attractor if $m>1$, since the eigenvalues of the corresponding linearization matrix are: $$\lambda_1=-\frac{3m+1}{m-1},\;\;\lambda_2=-\frac{2}{m-1}\;.$$ The NLED-dominated solution can be also a past attractor if $-1/3<m<1$. Otherwise it is a saddle point in $\Psi_{dgp}^-\cap\Psi_{dgp}^+$.

\end{itemize} while a similar couple of equilibrium points are found in $\Psi_{dgp}^-$ -- the phase space corresponding to the self-accelerating DGP brane:

\begin{itemize}

\item{Self-accelerating solution:} $$P_{self-a}^1=(0,0),\;\;\Omega_{cdm}=0,\;\;\Omega_B=0,\;\;q=-1\;,$$ which linearization matrix has the following eigenvalues: $$\lambda_1=-\frac{3}{2},\;\;\lambda_2=\frac{3m+1}{m-1}\;.$$

\item{Alternative self-accelerating solution:} $$P_{self-a}^2=(1,0),\;\;\Omega_{cdm}=0,\;\;\Omega_B=0,\;\;q=-1\;,$$ whose Jacobian matrix has the eigenvalues: $$\lambda_1=-\frac{3m+1}{m-1},\;\;\lambda_2=\frac{2}{m-1}\;.$$

\end{itemize} Both self-accelerating solutions are inflationary and, since $v=0$, they correspond to the late-time de Sitter phase, when $Q=0\;\Rightarrow\;H=1/r_c$. The first self-accelerating equilibrium point $P_{self-a}^1$ is the late-time attractor if $-1/3<m<1$. Otherwise ($m<-1/3$, $m>1$) it is a saddle equilibrium point in $\Psi_{dgp}^-$. Alternatively, the second self-accelerating critical point $P_{self-a}^2$ can be either, i) a saddle equilibrium point if $m>-1/3$, or ii) the future attractor if $m<-1/3$. This means that, for $m<1$, there always coexist two de Sitter (self-accelerating) solutions, one being a saddle equilibrium point, while the other one is the late-time (future) attractor. For $m>1$ the self-accelerating solution is a saddle critical point.

\subsubsection{Combination of positive and negative powers of $F$}

The next step is explore the effects of the DGP brane by considering the concrete Lagrangean (\ref{L}), which consists of positive and negative powers of the electromagnetic invariant $F$. The FRW field equations are:

\bea &&3Q_\pm^2=\left(\rho_{cdm}+\rho_B\right),\;\;Q_\pm^2\equiv H^2\pm\frac{H}{r_c}\nonumber\\
&&\frac{\dot Q}{Q}=-\frac{1}{2}H\left(\rho_{cdm}+\rho_B+p_B\right),\nonumber\\
&&\dot\rho_m=-3H\rho_{cdm},\;\;\dot B+2H B=0,\label{dgp feqs'}\eea where, as before, we are considering CDM and NLED-magnetic fluids on the brane, and $\rho_B$, $p_B$ are given by equations (\ref{rhop}). If we introduce the phase space variables:

\be x=\frac{\rho_B}{3Q^2},\;\;y=\frac{16\alpha B^4}{3Q^2},\;\;z=\frac{4\beta}{3Q^2B^2},\;\;v=\frac{Q}{H},\label{xyzv'}\ee then, the field equations (\ref{dgp feqs'}) can be written in the form of the following system of ODE:

\bea &&x'=(x-1)(x-y+z),\nonumber\\
&&y'=y(-5+x-y+z),\nonumber\\
&&z'=z(7+x-y+z),\nonumber\\
&&v'=-\frac{v}{2}\left(\frac{1-v^2}{1+v^2}\right)(3+x-y+z).\label{dgp asode'}\eea As before $$\Omega_{cdm}=v^2(1-x),\;\;\Omega_B=v^2x\;,$$ and, additionally: $$q=-1+\left(\frac{v^2}{1+v^2}\right)(3+x-y+z)\;,$$ and now the NLED-field EOS parameter does indeed depend on the equilibrium point under consideration: $$\omega_B=\frac{x-y+z}{3x}\;.$$ According to whether branch of the DGP model one is being considering, the phase space can be defined as 

\bea &&\Psi_{dgp}^-=\{(x,y,z,v)|0\leq x\leq 1,\;(y,z)\in\Re^2,\nonumber\\
&&\;\;\;\;\;\;\;\;\;\;\;\;\;\;\;\;\;\;\;\;\;\;\;\;8x+2y+z\geq 0,\;0\leq v\leq 1\},\label{phi-}\eea for the self-accelerating branch, or

\bea &&\Psi_{dgp}^+=\{(x,y,z,v)|0\leq x\leq 1,\;(y,z)\in\Re^2,\nonumber\\
&&\;\;\;\;\;\;\;\;\;\;\;\;\;\;\;\;\;\;\;\;\;\;\;\;\;\;\;\;\;\;8x+2y+z\geq 0,\;v\geq 1\},\label{phi+}\eea for the Minkowski cosmological phase instead. As before, due to the symmetry of the field equations (\ref{dgp feqs'}) under the replacement $Q\rightarrow-Q$, critical points with negative values of the $v$-s ($v<0$), are not being considered. 

The equilibrium points of (\ref{dgp asode'}) in the interception subset $\Psi_{dgp}^-\cap\Psi_{dgp}^+$ (GR-boundary), are:

\begin{enumerate}

\item{CDM-dominated solution:} $$P_{cdm}^{GR}=(0,0,0,1),\;\;\Omega_{cdm}=1,\;\;\Omega_B=0,\;\;q=\frac{1}{2}\;,$$ with the following eigenvalues of the Jacobian matrix $\lambda_1=3/2$, $\lambda_2=7$, $\lambda_3=-1$, and $\lambda_4=-5$. Hence, as before, this point is a saddle in the GR-boundary.

\item{Radiation-dominated phase:} $$P_{rad}^{GR}=(1,0,0,1),\;\;\Omega_{cdm}=0,\;\;\Omega_B=1,\;\;q=\frac{1}{2}\;.$$ The radiation-dominated solution ($\omega_B=1/3$), is also a saddle in the phase space, since the eigenvalues of the linearization matrix are of opposite signature: $\lambda_1=1$, $\lambda_2=2$, $\lambda_3=8$, $\lambda_4=-4$.

\item{UV NLED-dominated solution:} $$P_{nled}^{UV}=(1,-4,0,1),\;\;\Omega_{cdm}=0,\;\;\Omega_B=1,\;\;q=3\;.$$ Here, as in the GR case, the super-stiff ($\omega_B=5/3$), decelerating solution, is the past attractor, since $\lambda_1=\lambda_2=4$, $\lambda_3=5$, $\lambda_4=12$. It is associated with a big-bang cosmological singularity ($B^2/H^2=0\;\Rightarrow\;H\rightarrow\infty$).

\item{IR NLED-dominated phase:} $$P_{nled}^{IR}=(1,0,-8,1),\;\;\Omega_{cdm}=0,\;\;\Omega_B=1,\;\;q=-3\;.$$ This is the super-accelerating (phantom-like) solution, possibly associated with a big-rip-type singularity, as in the standard GR case. As in the pure GR situation, this equilibrium point is the late-time attractor since: $\lambda_1=-2,\;\;\lambda_2=-7$, $\lambda_3=-8$, $\lambda_4=-12$.

\end{enumerate} 

Meanwhile, in the phase space $\Psi_{dgp}^-$, corresponding to the self-accelerating branch of the DGP, one founds the following critical points:

\begin{enumerate}

\item{Self-accelerating phase 1:} $$P_{self-a}^1=(0,0,0,0),\;\;\Omega_{cdm}=\Omega_B=0,\;\;q=-1\;.$$ The eigenvalues are: $$\lambda_1=7,\;\;\lambda_2=-3/2,\;\;\lambda_3=-1,\,\;\lambda_4=-5\;.$$

\item{Self-accelerating phase 2:} $$P_{self-a}^2=(1,0,0,0),\;\;\Omega_{cdm}=\Omega_B=0,\;\;q=-1\;,$$ with the following eigenvalues: $$\lambda_1=1,\;\;\lambda_2=8,\;\;\lambda_3=-2,\,\;\lambda_4=-4\;.$$

\item{Self-accelerating phase 3:} $$P_{self-a}^3=(1,0,-8,0),\;\;\Omega_{cdm}=\Omega_B=0,\;\;q=-1\;.$$ Here, $$\lambda_1=2,\;\;\lambda_2=-7,\;\;\lambda_3=-8,\,\;\lambda_4=-12\;.$$

\item{Self-accelerating phase 4:} $$P_{self-a}^4=(1,-4,0,0),\;\;\Omega_{cdm}=\Omega_B=0,\;\;q=-1\;.$$ The eigenvalues of the linearization matrix are: $$\lambda_1=4,\;\;\lambda_2=5,\;\;\lambda_3=12,\,\;\lambda_4=-4\;.$$

\end{enumerate} These four critical points correspond to the inflationary, late-time self-accelerating solution $H=1/r_c$ ($v=0\;\Rightarrow\;Q=0$), and are saddle equilibrium points in $\Psi_{dgp}^-$.

As seen, the Minkowski cosmological phase of the DGP does not modify the stability properties of the GR equilibrium points found in section IV. The self-accelerating branch of the DGP differs from results in that section in the addition of four critical points, associated with the self-accelerating de Sitter solution $H=1/r_c$.

\section{Discussion and Conclusions}

Non-linear electrodynamics can supply a useful scenario where to discuss such relevant problems of the standard cosmological model as the initial
singularity related with the big-bang, and the early-times and present stages of accelerated expansion of the universe. A Lagrangean density of the form given in (\ref{L}), for instance, can account for the present accelerated stage of the cosmic evolution, as well as for avoidance of the initial singularity, in a unified picture \cite{novellocqg}. In a general setting, where arbitrary powers of $F$ are invoked, this behavior can be explained as due to the asymptotic properties inherent in these kinds of Lagrangean. Actually, given a Lagrangean of the form $$L=L_0 F^{1/(1-m)}\;,$$ it has been shown in section III, that, for $m>1$, the matter-dominated solution is the past attractor, while the NLED inflationary solution ($\omega_B<0,\;q<0$) is the future/late-time attractor. For $m<1$, these solutions exchange their stability properties: the NLED non-inflationary solution ($\omega_B>0,\;q>0$) is the past attractor, while the matter-dominated solution (also non-inflationary since $q=1/2$), is the late-time attractor. While none of these models alone seems like a plausible cosmological model, combinations of negative and positive powers of the electromagnetic invariant $F$, provide an interesting cosmological scenario instead.

To make the analysis simpler, we have found instructive to consider the so called ''magnetic universe'' (no electric component). For definiteness we have explored a FRW universe whose dynamics is based on the Lagrangean $$L_{tot}=L_g+L_{cdm}+L(F)\;,$$ where $L_g$ is the standard Einstein-Hilbert Lagrangean, $L_{cdm}$ corresponds to a pressureless perfect fluid (CDM), and $L(F)$ is the NLED Lagrangean displayed in Eq. (\ref{L}). By means of the dynamical systems tools we are able to find the relevant asymptotic dynamics of the model. The goal is to write the original system of cosmological equations in the form of an autonomous system of ordinary differential equations. Then, the information on the dynamics is encoded in the structure of the phase space, in particular in the knowledge of the equilibrium/critical points and their stability properties. The knowledge of the equilibrium points in the phase space corresponding to a given cosmological model is a very important information since, independent on the initial conditions chosen, the orbits of the corresponding autonomous system of ODE will always evolve for some time in the neighborhood of these points. Besides, if the point were a stable attractor, independent on the initial conditions, the orbits will always be attracted towards this point (either into the past or into the future). Going back to the original cosmological model, the existence of the equilibrium points can be correlated with generic cosmological solutions that might really deside the fate and/or the origin of the cosmic evolution. In a sense the knowledge of the asymptotic properties of a given cosmological model is more relevant than the knowledge of a particular  analytic solution of the corresponding cosmological equations. While in the later case one might evolve the model from given initial data giving a concrete picture that can be tested against existing observational data, the knowledge of the asymptotic properties of the model gives the power to realize which will be the generic behaviour of the model without solving the Einstein's field equations. In the dynamical systems language, for instance, a given particular solution of the Einstein's equations is just a single point in the phase space. Hence, phase space orbits show the way the model drives the cosmological evolution from one particular solution into another one. Equilibrium points in the phase space will correspond to solutions of the cosmological (Einstein's) equations that, in a sense, are preferred by the model, i. e., are generic. The lack of equilibrium points that could be correlated with a given analytic solution of the model, amounts to say that this solution is not quite generic, otherwise unstable in terms of phase space variables, and can not be taken too seriously.

The asymptotic structure of the FRW cosmological model based on the above Lagrangean $L_{tot}$ shows that the UV NLED-dominated, super-stiff-fluid (decelerating expansion) solution, is the starting point of the cosmic evolution (all orbits in the phase space emerge from this solution). The radiation and CDM-dominated solutions are also critical points of the corresponding system of ODE, so that there is room in this model to explain the observed amount of structure we see. Besides, the IR NLED-dominated, super-inflationary solution is the future/late-time attractor, meaning that it is the end-point of the cosmic evolution. Due to the term $\propto\beta B^{-2}$ in the NLED Lagrangean (\ref{L}), the NLED-fluid at late times behaves like phantom matter: the IR NLED-dominated future attractor could be associated with a big-rip type of cosmological singularity. Hence, in the above model both, the starting singular point of the cosmic evolution -- the bib bang singularity --, as well as the catastrophic fate of the cosmic evolution, are unavoidable. The fact that the past attractor is associated with the initial cosmological singularity means that, contrary to what former authors found \cite{16,novellocqg}, avoidance of the big bang singularity is not a generic property of models of the kind given by the Lagrangean (\ref{L}). Only under very specific initial conditions, or under very special considerations, solutions were the singular origin of the evolution is evaded, may be found.

When the brane effects are taken into consideration, one finds that while the DGP brane does not appreciably affects the cosmic dynamics driven by the NLED Lagrangean (\ref{L}) -- but for the occurrence of critical points associated with the de Sitter solution $H=1/r_c$ in the self-accelerating branch of the DGP model --, the Randall-Sundrum brane effects do modify the stability properties of the GR equilibrium points. The interesting finding is that the extra-dimensional effects modify not only the UV-asymptotics, but also the IR/late-times dynamics. Actually, once the RS2 brane effects are considered, the attractor structure of the UV (past attractor) and IR (future attractor) NLED-dominated solutions, within GR, is demoted to just saddle property. In other words, the consequence of the extra-dimensional (RS2) brane effects is that, both NLED-dominated critical points above, represent just saddle equilibrium points in the phase space. This would mean, in turn, that the big-bang (initial) and the big-rip (final) cosmological singularities, can be evaded just by a proper choice of the initial conditions: there are orbits of the autonomous system of ODE that do not pass through the neighborhood of the corresponding critical points.

The finding that the stability properties of the infrared (IR) NLED-dominated solution are modified by the RS2 brane effects, came as a surprise. Actually, only at very high energies can the graviton acquire large momenta along the extra dimension and may escape into the bulk (5D) spacetime. On the contrary, at low energies (large cosmological scales), it is expected that the RS2 brane effects can be safely ignored. However, while making such statements one has to be careful. In the cosmological context, the most evident RS2 brane effect is to modify the Friedmann equation: $$3H^2=\rho_T\left(1+\frac{\rho_T}{2\lambda}\right)\;.$$ Hence, at very high energy density $\rho_T\gg\lambda$ (much bigger than the brane tension), the Friedmann equation is fundamentally modified $3H^2\propto\rho_T^2$. If in the course of the cosmic expansion the total energy content of the universe dilutes, then as long as $\rho_T\ll\lambda$, one recovers standard GR-Friedmann behavior. Now look at the Lagrangean density for the NLED-magnetic field (\ref{L}) considered here. Note that as the expansion proceeds the magnetic field $F\propto B^2$ dilutes, and the component $\propto\beta B^{-2}$ in (\ref{L}) grows without limit. This means that the total energy content of the universe starts growing at the expense of the NLED component so that, at late times, eventually, $\rho_T$ might become much larger than the brane tension once again, rendering the brane effects important at late times also. This effect has been demonstrated to be generic of phantom fields of any nature which are trapped in a RS2 braneworld \cite{companion}. Here we have demonstrated that the above mixing of UV/IR scales, due to extra-dimensional effects on the dynamics of phantom matter, is distinctive only of models where the brane effects modify the right-hand-side (energy density) of the Friedmann equation, so that, for instance, the DGP brane model does not show this mixing.

Summarizing: we have investigated the asymptotic structure of magnetic-NLED models where combinations of positive and negative powers of the electromagnetic invariant $F$ are considered. It has been demonstrated that such models can supply an interesting cosmological scenario where the end-point of the cosmic evolution is a super-accelerated, singular state indistinguishable from the big-rip inherent in (scalar) phantom driven cosmologies. RS brane effects may drastically change the nature of the starting point, as well as the fate of the cosmic evolution: both, the big bang singularity, and the big rip event, can be avoided. Perhaps one of the most unexpected findings of the present research, was to show that the unusual behavior of the energy density associated with phantom-like matter (mimicked here by an appropriate NLED Lagrangean), in conjunction with the Randall-Sundrum brane effects, results in a model where the laws of gravity are simultaneously modified in the UV and in the IR limits (UV/IR mixing of scales). The above mixing is distinctive only of models that modify the matter part of thee Friedmann equation. 

This work was partly supported by CONACyT M\'{e}xico, under grants 49924-J, 105079, and by Instituto Avanzado de Cosmologia (IAC) collaboration. R. G.-S. acknowledges partial support from COFAA-IPN and EDI-IPN grants, and SIP-IPN 20100610. T. G. acknowledges also the MES of Cuba for partial support of the research.

\section{Appendix: Dynamical Systems}

Here we include brief tips of how to apply the dynamical systems tools to situations of cosmological interest. In order to apply these tools one has to follow the steps enumerated below: 

\begin{enumerate}

\item To identify the phase space variables that allow writing the system of cosmological equations in the form of an autonomus system of ordinary differential equations (ODE), say:\footnote{There can be several different possible choices, however, not all of them allow for the minimum possible dimensionality of the phase space.}$$x_i=(x_1,x_2,...x_n)\;.$$

\item With the help of the chosen phase space variables, to build an autonomous system of ODE out of the original system of cosmological equations ($\tau$ is the time-ordering variable, not necessarily the cosmic time): $$\frac{dx_i}{d\tau}=f_i(x_1,x_2,...x_n)\;.$$ Notice that the RHS of these equations do not depend explicitly on $\tau$ (that is the reason why the system is called autonomous).

\item To identify the phase space spanned by the chosen variables $(x_1,x_2,...x_n)$, that is relevant to the cosmological model under study. This amounts, basically, to define the range of the phase space variables that is appropriate to the problem at hand: $$\Psi=\{(x_1,x_2,...x_n):\text{bounds on the}\;x_i\text{-s}\}\;.$$ 

\item Finding the equilibrium points of the autonomous system of ODE, amounts to solve the following system of algebraic equations on $(x_1,x_2,...x_n)$: $$f_i(x_1,x_2,...x_n)=0\;.$$

\item Next one linearly expands the equations of the autonomous system of ODE in the neighborhood of the equilibrium points $\bar p_k=p_k(\bar x_1,\bar x_2,...\bar x_n)$, $k=1,2,...m$:\footnote{In general the number of equilibrium points is different from the dimension of the phase space: $m\neq n$.} I. e., one replaces $x_i\rightarrow\bar x_i+e_i$, where $e_i$ are the small (linear) perturbations around the equilibrium points. Hence the system of ODE becomes a system of linear equations to determine the evolution of the $e_i$-s: $$\frac{de_i}{d\tau}=\bar f_i+\sum_{j=1}^n\left(\frac{\partial f_i}{\partial x_j}\right)_{\bar p}e_j+{\cal O}(e_i^2)\;,$$ otherwise, since $\bar f_i=f_i(\bar p)=0$, then $$\frac{de_i}{d\tau}=\sum_j^n[M(\bar p)_i^j]\;e_j+{\cal O}(e_i^2)\;,$$ where we have introduced the linearization or Jacobian matrix $[M^j_i]=\partial f_i/\partial x_j$.

\item The next step is to solve the secular equation to determine the eigenvalues $\lambda_i$ of the linearization matrix at the given equilibrium point $\bar p$: $$\det|M(\bar p)^j_i-\lambda\;U^j_i|=0\;,$$ where $[U^j_i]$ is the unit matrix. 

\item Once the eigenvalues of the linearization around a given equilibrium point $\bar p$ have been computed, the evolution of the perturbations is given by $$e_i(\tau)=\sum_j^n (e_0)_i^j\exp{(\lambda_j\tau)}\;,$$ where the amplitudes $(e_0)_i^j$ are constants of integration.

\end{enumerate} If all of the eigenvalues have negative real parts, the perturbations decay with $\tau$, i. e., the equilibrium point is stable against linear perturabtions. The corresponding equilibrium point is said to be a future attractor. If at least one of the eigenvalues has positive real part, the perturbations grow with $\tau$ so that these are not stable in the direction spanned by the given eigenvalue. Hence the point is said to be a saddle. The perturbations around a given equilibrium point are unstable, in other words the point is a past attractor -- also acknowledged as a source point in the phase space --, if all of the eigenvalues have positive real parts. Points whose linearization is characterized by complex eigenvalues are said to be spiral equilibrium points, and are commonly associated with oscillatory behavior of the corresponding solution. If at least one of the eigenvalues has a vanishing real part, the equilibrium point is said to be non-hyperbolic. In the latter case, in general, and unless some of the non-vanishing real parts of the eigenvalues are of opposite sign, one can not give conclusive arguments on the stability of the equilibrium point. Other techniques have to be applied.

\end{document}